# Three Layer Hierarchical Model for Chord

Waqas A. Imtiaz, Shimul Shil, A.K.M Mahfuzur Rahman

*Abstract*— Increasing popularity of decentralized Peer-to-Peer (P2P) architecture emphasizes on the need to come across an overlay structure that can provide efficient content discovery mechanism, accommodate high churn rate and adapt to failures in the presence of heterogeneity among the peers. Traditional p2p systems incorporate distributed client-server communication, which finds the peer efficiently that store a desires data item, with minimum delay and reduced overhead. However traditional models are not able to solve the problems relating scalability and high churn rates. Hierarchical model were introduced to provide better fault isolation, effective bandwidth utilization, a superior adaptation to the underlying physical network and a reduction of the lookup path length as additional advantages. It is more efficient and easier to manage than traditional p2p networks. This paper discusses a further step in p2p hierarchy via 3-layers hierarchical model with distributed database architecture in different layer, each of which is connected through its root. The peers are divided into three categories according to their physical stability and strength. They are Ultra Super-peer, Super-peer and Ordinary Peer and we assign these peers to first, second and third level of hierarchy respectively. Peers in a group in lower layer have their own local database which hold as associated super-peer in middle layer and access the database among the peers through user queries. In our 3-layer hierarchical model for DHT algorithms, we used an advanced Chord algorithm with optimized finger table which can remove the redundant entry in the finger table in upper layer that influences the system to reduce the lookup latency. Our research work finally resulted that our model really provides faster search since the network lookup latency is decreased by reducing the number of hops. The peers in such network then can contribute with improve functionality and can perform well in P2P networks.

*Keywords- Hierarchy; DHT; CHORD; P2P.*

## I. INTRODUCTION

Peer-to-Peer (P2P) network is a logical overlay network which is built on top of one or more existing physical networks. P2P networks, over the last two decades has been recognized as a more efficient and flexible approach for sharing resources, compared to the traditional Client-Server model. Internet-scale decentralized architecture bases on p2p, created an environment for millions of users, allowing them to simultaneously connect and share content with ease and reliability [1][8]. Efficient data location, lookups, redundant storage and distributed content placement of p2p overlay networks have raised a big deal of attention, not only for researchers/academicians but also in practical usage of the technology [7]. The distributed approach of p2p system is less vulnerable to attacks, robust and highly available as compared to its client-server counterpart.

P2P algorithms are a class of decentralized distributed systems collectively, called as Distributed Hash Tables (DHTs). DHT is a distributed data structure whose main function is to hold the key-value pair in a completely distributed manner and any participating peer in the overlay network can retrieve the value associated with the given key [8]. DHT uses consistent hashing to map the responsible node for a key-value pair. Along with efficient mapping of a key-value pair to nodes, DHT also has the ability to isolate the network changes to a small part of it thus limiting the overhead and bandwidth consumption during networks and resources updates [8]. DHT is an abstract idea that helps us to achieve complete independence from a central lookup entity and tolerance to changes in the network [9]. Different algorithms have been designed to implement and refine DHT idea. CAN, PASTRY, TAPSTERY, CHORD are the most popular implementations of DHT. We choose chord because of well-organized data structures and efficient routing schemes.

Chord is an efficient distributed lookup service based on consistent hashing which provides support for only one operation: given a key and it efficiently maps the key onto a node [7]. At its heart chord provides a fast and efficient computation of the hash function by mapping keys onto the nodes [7]. Chord basically creates a one dimensional identifier circle which ranges from 0 to $(2^m - 1)$, where m is the number of bits on the identifier circle and every node and key on the identifier circle are assigned an m-bit identifier. Node identifier is obtained by hashing the port number and node's IP address, whereas key identifier, 160 bits, is created by hashing the key [7]. Identifiers are assigned in such a way that the probability of two nodes having same hashed identifiers is negligible [7].

The main advantage of Chord in a large-scale setting is its high scalability and self-organization. But with the large amount of users, high churn condition, introduce a high overhead while maintaining the DHT structure. Further, dynamic joining and leaving of nodes may results in loss of key-value pairs, even though their respective nodes are still alive. Controlling this situation is difficult and incurs more overhead. While majority of the peers are short lived and have minimal capabilities, a small percentage typically remains up for long periods and have relatively better storage, bandwidth and memory. This property has been used to design hierarchical models, where more stable nodes can dynamically form an upper level overlay. Other short lived peers can get themselves connected as sub-overlay of these upper level nodes.

Naturally hierarchical DHT design has a better fault isolation, effective bandwidth utilization, superior adaptation to the underlying physical network and a reduction of the lookup path length as an additional advantage [14]. It is more efficient and easier to manage than the pure p2p structure. Moreover by dividing the whole system into several different layers that tries to solve local tasks inside their own layers, results in reduced workload and efficient operation of the network [48].

Keeping in view the advantages of hierarchal models over traditional p2p models, this paper proposes a three layer DHT based overlay, with decentralized key database architecture that relies on p2p algorithm Chord. This projected three layer p2p architecture with decentralized database consists of a number of database subsystems, each of which is connected to others through its root only.

To meet the challenges of present internet applications, this paper presents a three hierarchical model presented in [1]. The lower layer peers are designed to be deployed on resource constraints. In p2p overlay network lower layer peers do not need to deliver high data rates and high computational power. The second layer act as a medium between lower and upper layer. Middle layer is designed by the super peer. This super peer performs as a central server for lower layer peers. The upper layer is the core or backbone layer in this architecture. We design this layer with ring based ultra-superpeer communication. Ultra-superpeer in upper layer also acts as a central server for associated middle layer super-peers. Finally, each layer super -peers perform as a central server for immediate lower layer peers. Each ultra-superpeer maintains the pointer index and a finger table that contains the successor/predecessor list in chord ring and index contains the all previous level peer ID and data list, but does not store the data or document. Moreover, super-peer in the middle layer only maintains the pointer index. Each of the ordinary peers is attached to a super peer with point-to-point or mesh topology connection and this super peer belongs to both the ordinary peers and the super-peers. As a result local peers do not have to share the burden of possibly high maintenance traffic and the overlay network does not have to deal with their performance bottlenecks and low reliability [7].

## II. SYSTEM MODEL

Based on the capacity and availability of peers in the overlay network, they are classified into three categories: Ordinary peers, Super-peers and Ultra Super-peers as shown in fig. 1. Each peer in the overlay network has a peer ID, which is computed from its IP address and port address pair using consistent hash function (SHA-1) [10]. Distributed data sharing techniques are applied in the hierarchical p2p model. A data item is represented by a key-value pair. Where key is the data name and value is the content associated with the key [10]. A global consistent hash function is used to map object key to peer identifiers. A peer uses a *put (key, value)* command to insert the data item and *get (key)* command for obtaining the value of the data item in the overlay [10]. In the 3-layer hierarchical based p2p system, if the lower layer ordinary peer wants to share the file, it is necessary to send the metadata information to its associated superpeer in the middle layer.

The superpeer keeps the metadata in the index. As shown in fig. 2, superpeer index has two field, key_ID and peer_ID. Where key_ID is the file sharing identity and peer_ID is the ordinary peer identity. However, it is necessary to further declare the metadata information to the associated ultra-superpeer in extended format that are kept in the index of the ultra-superpeer. The extended index has three field, key_ID, peer_ID and superpeer_ID as shown in fig. 3. The superpeer_ID is the associated superpeer of the ordinary peer. At the query process, the metadata index is to provide the sufficient information [15].

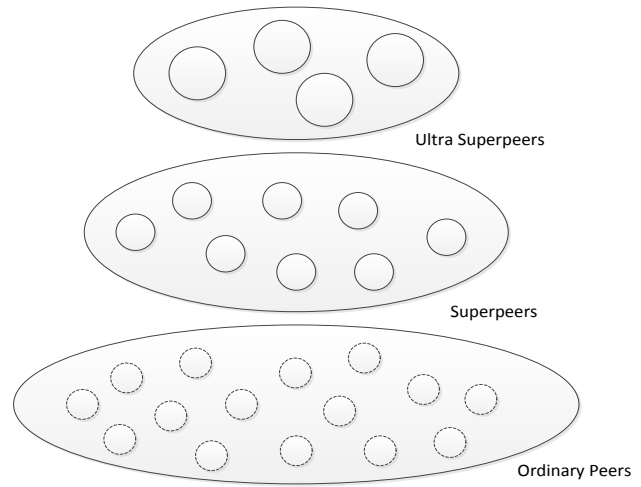

Fig 1: Three layer hierarchical Model for Chord

| Key_ID | Peer_ID |
|--------|---------|
| X | N12 |
| ……… | ……… |
| ……… | ……… |

| Key_ID | Peer_ID | Superpeer_ID |
|--------|---------|--------------|
| X | N12 | N1 |
| …………… | …………… | …………… |
| …………… | …………… | …………… |
| …………… | …………… | …………… |

Fig 3: Ultra Superpeer Index

### A. Ordinary Peers

The ordinary peers form the single connection structure where peers communicate only with their super-peers [14]. All peers frequently join and leave the lower layer without affecting the entire overlay network. Ordinary peers are the members of the lower layer hierarchy and may disconnect after sharing their resources over p2p network. Ordinary peer can communicate with other ordinary peers through associated super-peers in the second layer.

In the single-connection structure, every ordinary peer uses the PING/PONG algorithm periodically to check the connectivity with super-peer. The ordinary peers send the PING message to its super-peer and the super-peer responds with a PONG message [7] [14]. As a result, at least two hops are needed to communicate with each other [13] [14].

Ordinary peers become a super-peer by sending an upgrade (capability and availability) level request to the associated super-peers. The current super-peer creates a list of ordinary peer request which may become a super-peer after leaving the current super-peer [11]. Ordinary peers ID in the lower layer groups must be smaller than the associated super-peer ID.

*B. Super Peers*

Peers with more stability and storage are placed in the middle layer of hierarchical system. Super-peers are also organized by the single connection structure. So that, middle layer super-peers check their connectivity with the upper layer via PING/ PONG message. Moreover, super-peers act as a server to maintain the index pointer information of ordinary peers and are also responsible for the query an object in the overlay network [7] [13].

In our hierarchical architecture, peers ID in the lower layer groups must be smaller than the associated super-peer ID and the super-peer ID in the middle layer must lay between the ultra-superpeer ID and predecessor ID in the upper layer. Every super-peer contains the metadata of the lower layer ordinary peers as index pointer.

*C. Ultra Super peers*

Ultra super-peer are peers with ideal characteristic, higher availability and which are being predicted to be available in future for a long period compared to the middle layer super-peer of the hierarchical structure. Ultra super-peers are organized in a circle where peers are connected to one another similar to that of a chord network. Moreover, ultra super-peer is the gateway or path to communicate among the super-peers in the middle layer, and acts as a central server for associated middle layer super-peers. In the chord ring each ultra-superpeers apply the *stabilization* protocol periodically to update the successor / predecessor pointer and the optimized finger table in the event of peer failure or migration from middle layer [7] [13] [14].

III. NODES PLACEMENT

Proposed three layer hierarchical model, applies different methods for joining and leaving of peers in overlay network.

*A. Peers Joining*

Peer that joins that overlay network, will become an ordinary peer. Ordinary peer cannot directly opt of becoming a superpeer or an ultra-superpeer. Ordinary peers can only migrate to the immediate upper layer based on their physical strengths i.e. bandwidth, stability, storage space etc.

The joining peer using any bootstrapping method sends a joining request to an existing known superpeer. Superpeer examines the peer ID of the joining peer and finds out a place for the joining peer in the overlay network [7]. After finding a place, superpeer sends a response message to the joining peer with the information of joining place. Ordinary peer on receiving the necessary information joins the overlay network and sends the metadata to its associated superpeer. Corresponding ultra superpeer is also informed of the new entry by its associated superpeer [11] [12].

*B. Peers Leaving*

Peers are classified into three layers that comprise the hierarchical model. Peers from each layer can leave the overlay network in the following manner:

 *1) Ordinary Peer Leaves:* Ordinary peers may leave the hierarchy without informing any other peers because it does not maintain the routing table. Superpeer detects the leaving of an ordinary peer by PING/PONG message in the lower layer. When superpeer detects a leaving ordinary peer, it changes the status of the leaving peer by deleting its associated metadata. Superpeer also informs the associated ultra superpeer in upper layer to delete the metadata of the corresponding ordinary peer.

 *2) Super Peer Leaves:* As superpeer is responsible for multiple tasks in the hierarchy, so it cannot leave the system without any prior arrangements. Before superpeer leaves the system, it must select a candidate superpeer from the ordinary peers in the group.
After selecting a new superpeer, all metadata is transferred to the new superpeer. The new superpeer migrates to the middle layer and operates in place of the leaving superpeer. Migrated candidate superpeer also informs its connected ultra-superpeer to update the metadata and its associated ordinary peers about its arrival.

 *3) Ultra Super Peer Leaves:* Ultra-superpeer selection and migration in the overlay network is similar to the superpeers leaving process, except the backup of metadata. Ultra-superpeers use the DHT chord protocol to periodically updates itself and uses its successor from finger table as backup of the metadata. The new ultra-superpeer on joining the upper layers, also informs its successor. When the candidate ultra-superpeer migrates to the upper layer, at the same time candidate superpeer also migrates to the middle layer to fix the link among the three layers. It is exceptional to leaves the ultra superpeer because of its stability.

*C. Peers Migration*

When a superpeer in the middle layer fails or leaves the overlay network, a new superpeer will be needed to establish communication between the associated ordinary peers and to perform the query process [7]. Ordinary peers previously marks as candidate superpeers are selected to replace the failed superpeer [1]. If the system cannot find any candidate superpeer, then an ordinary peer is selected on the basis of capacity, firewall support and availability. After finding the candidate Superpeer, it is migrated to the middle layer in place of the leaving Superpeer. Same procedure is also applied between the middle and upper layer in hierarchy, if the ultra superpeer fails or leaves. As a result, the number of Superpeers in the middle layer and ultra-superpeers in the upper layer is possibly unchanged.

IV. LOOKUP PROCESS

Lookup process of the proposed 3 layer hierarchy is shown in fig. 4. Lookup process starts at the ordinary peer, where data ID of a particular data item is needed to look up the peers. The data ID is obtained by hashing the data key [10]. Query message containing the required data ID is forwarded from the ordinary peer to the associated superpeer. After getting the query request, superpeer checks its database against the provided data ID. If the data ID is found in the ordinary peer groups, lookup process is done, otherwise the query request is forwarded to the connected ultra-superpeer in the ring [1][14]. When the query request arrives at associated ultra- superpeer, it checks its own database. If data item is not found, it forwards the query to its successor based on the finger table similar to chord ring, and tries to find the data item. Ultra-superpeers in upper layer collectively contain all the overlay peers metadata information. So that data item is found efficiently, if it is available in the overlay.

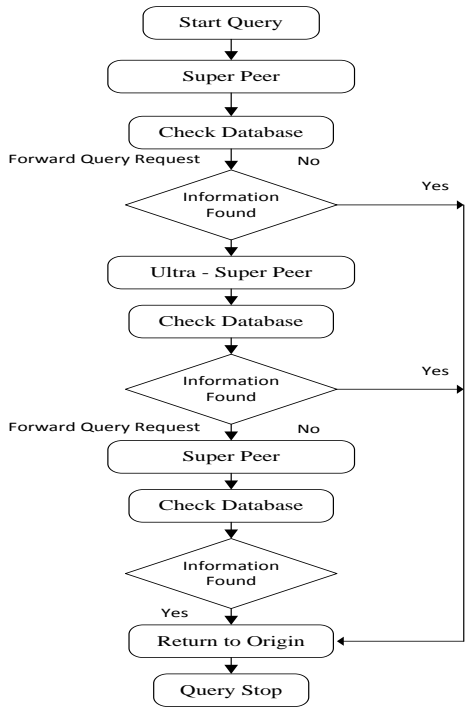

Fig 4: Hierarchical Model lookup process

## V. ANALYTICAL ANALYSIS

Performance of the proposed system can be measured on the basis of lookup latency and hoops per lookup. We compare conventional chord with our proposed three layer hierarchical chord to identify the efficient model for p2p applications.

For conventional chord we have assumed the following parameters [11][14].

- Peer and key identifiers are random.
- Number of peers = N
- Number of entries in Finger table = $O(\log N)$
- Probability of the number of forwarding = $O(\log N)$

Because of the stable finger table in conventional chord, the average hops required to complete a lookup process is $\frac{1}{2} \log_2 N$ [1], where N is the number of peers in the overlay network. For our proposed three layer hierarchical model, we assume single connection structures, and take the following parameters for comparison [7][11][14][15]:

- Total number of peers in the overlay = N
- Number of ultra-superpeers = U
- Number of superpeers = S
- Number of Ordinary peers = $\{N - (S + U)\}$
- Number of peers in a lower layer group = $\frac{N-(S+U)}{S}$
- Probability of finding data item at super node = Q
- Number of entries in Finger table = $O(\log U)$
- Time complexity is $O(\log U + 1 + 1)$

Time complexity is the least amount of time required to execute a process or program. When the number of peers in the overlay network is N, the typical time complexity of searching is $O(N)$ for unstructured p2p network and $O(\log N)$ for structured p2p network [1]. We have used unstructured topology in the lower two layers and structure topology in the upper layer. In lower and middle layer, we apply the single connection intra group structure, therefore the typical time complexity of searching is $O(1)$, as single peer is traversed in order to complete the query the query process. In upper layer, the typical time complexity of searching is $O(\log U)$ because we apply structured based chord protocol. Finally, total time complexity of the model is $O(\log U + 1 + 1)$.

In our proposed hierarchy model, three cases occur during lookup process i.e. average number of hops required 1, if the query process is done between ordinary peers and superpeers. Average number of hops required is (1+1) = 2, if the query process is done among ordinary peers, super peers and associated ultra-superpeer (without using chord ring) because of single connection structure. Average number of hops required ½ $log$U, if the query process is done in upper layer chord ring. Thus total number of hops required to perform a query process in ½ $log$ (U+2).

Lookup latency considered in our model is half of the time required for a peer in the traditional chord. Because in the hierarchical model, ultra-superpeers and superpeers, performs the lookup process and have more physical capability then the ordinary peers. That is why we assume the average link latency of a peer in our hierarchical system is 50 % less than the traditional chord peer which are unstable and have minimum computation capabilities.

To analyze the performance of both traditional chord and our proposed three layer hierarchical model we randomly increase the number of nodes in the overlay network and observe the number of hops and time taken to perform the lookup process. Number of superpeers and ultra superpeers are also randomly increased while increasing the number of nodes.

Figure 5 represents the number of hops required to perform the lookup process against increasing number nodes. Blue line shows the number of hops required by traditional chord, whereas green line represents the number of hops used by three layered hierarchical chord to perform the lookup process. Figure shows that as we increase the number of nodes, the number of hops required to perform the query process also increases. However the rate of increase is abrupt in traditional chord as compared to our proposed model. This is because of the fact that traditional chord uses $\frac{1}{2}(\log N)$ hops for lookup, and as the number of nodes increases, the number of hops required to perform a lookup process also increases. Our hierarchical model uses $\frac{1}{2}(\log U + 2)$ hops which significantly reduces the number of hops required to perform lookups, as the only factor that can increase the number of hops is sufficiently small to produce any significant effect.

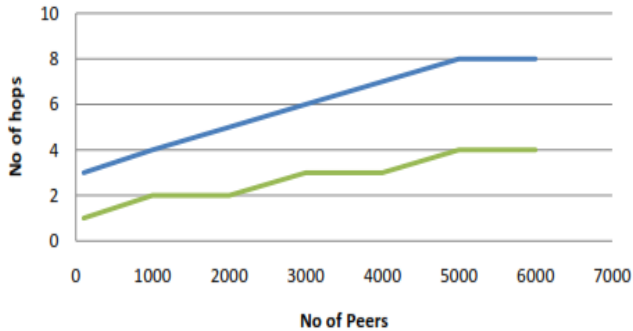

Fig 5: Number of Hops per lookup

Figure 6 shows the lookup latency against increasing number of nodes. It is obvious that as the number of nodes increases the time taken by traditional to perform lookups is greater than that of our hierarchical model. This is because the number of nodes required to perform lookup is less as shown in figure 4, as well as the time taken by each individual node in our hierarchical model is comparatively less that nodes in the traditional chord. This is why, when the number nodes increases, the time taken to perform lookups in traditional chord increases rapidly as compared to our model.

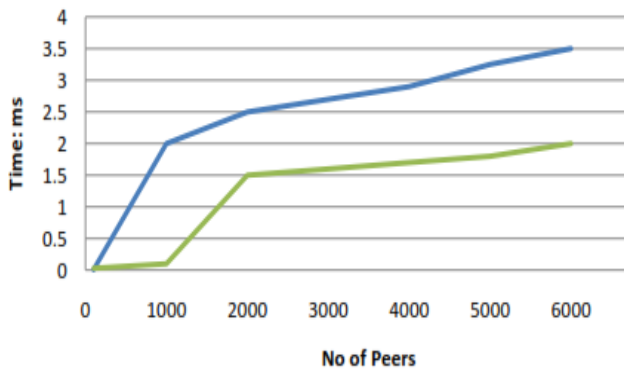

Fig 6: Lookup Latency

## VI. CONCLUSIONS

This paper presents a three layer hierarchical overlay architecture, which is efficient, stable, and scalable as shown. Peers in such architecture can frequently join or leave the network, whereas migrating to different layers are also presented in our proposed model. This model is able to discriminate among the peers according to their capability and thus classifying them into three categories: Ultra-Superpeer, Superpeer and Ordinary Peer.

Our hierarchical design offers higher stability by using ultra-superpeers at the upper layer which are more reliable peers. We also presented an instantiation of our 3-layer hierarchical model using Chord at the upper layer and superpeer based single connection intra group structure in middle layer. Since the number of Ultra-Superpeer is less than the number of Superpeer and the number of Superpeer is less than the ordinary peer, the overhead to find a desired data has become less than two layer architecture by keeping metadata at each layer. To reduce the access load of the database, we used decentralized database to distribute the query load. That approach gives a guarantee for the system to say that there will be no single point of failure. Superpeer and DHT chord based lookup process are used in the hierarchy which also helps to reduce the average number of hops to find the desired peer holding the desired data item.

Lookup latency of the model is also decreased by confirming that the average number of required hops is decreased to find the desired peer. This model executes the lookup process with the help of superpeer and chord protocol that extensively reduces the query traffic load which is common characteristic of the flooding based query. Overall structure ensures that database load is also minimized as the load of a particular peer is split among the neighbor peers. Proposed model offers more stability and scalability than existing chord algorithm, however a lot of work and analysis is required to implement and use this model in practice.